\documentclass[10pt,journal]{IEEEtran}

\usepackage{graphicx}
\usepackage{fixltx2e}
\usepackage{stfloats}
\begin{document}
%
% paper title
% can use linebreaks \\ within to get better formatting as desired
\title{Improving Recall of \textit{In Situ} Sequencing by Self-Learned Features and a Graphical Model}

% author names and affiliations
%
% If one affiliation:
%
%   \author{\IEEEauthorblockN{%
%   Homer Simpson,
%   James Kirk,
%   Montgomery Scott and
%   Eldon Tyrell}
%   \IEEEauthorblockA{Starfleet Academy, San Francisco, California 96678-2391\\
%   Telephone: (800) 555--1212, Fax: (888) 555--1212\\
%   Email: $\{$shell,simpson,kirk,scott,tyrell$\}$@starfleet.com}
%   }
%
% If multiple affiliations:
%
\author{\IEEEauthorblockN{%
Gabriele Partel\IEEEauthorrefmark{1},
Giorgia Milli\IEEEauthorrefmark{2} and
Carolina W{\"a}hlby\IEEEauthorrefmark{1}}\\
\IEEEauthorblockA{\IEEEauthorrefmark{1}Centre for Image Analysis, Uppsala University, Sweden.\\
Email: $\{$gabriele.partel,carolina.wahlby$\}$@it.uu.se}\\
\IEEEauthorblockA{\IEEEauthorrefmark{2}Politecnico di Torino, Italy.\\
Email: giorgia.milli@studenti.polito.it}
}

% make the title area
\maketitle

\begin{abstract}
Image-based sequencing of mRNA makes it possible to see where in a tissue sample a given gene is active, and thus discern large numbers of different cell types in parallel. This is crucial for gaining a better understanding of tissue development and disease such as cancer. Signals are collected over multiple staining and imaging cycles, and signal density together with noise makes signal decoding challenging. Previous approaches have led to low signal recall in efforts to maintain high sensitivity. We propose an approach where signal candidates are generously included, and true-signal probability at the cycle level is self-learned using a convolutional neural network. Signal candidates and probability predictions are thereafter fed into a graphical model searching for signal candidates across sequencing cycles. The graphical model combines intensity, probability and spatial distance to find optimal paths representing decoded signal sequences. We evaluate our approach in relation to state-of-the-art, and show that we increase recall by $27\%$  at maintained sensitivity. Furthermore, visual examination shows that most of the now correctly resolved signals were previously lost due to high signal density. Thus, the proposed approach has the potential to significantly improve further analysis of spatial statistics in \textit{in situ} sequencing experiments.

\end{abstract}

% Don't add keywords to the paper!

\section{Introduction}
% no \IEEEPARstart
\textit{In situ} sequencing \cite{ke2013situ} is a very powerful tool to quantify gene expression directly in biological tissue samples without losing spatial information on tissue morphology. Investigated genes are targeted with controlled design of barcoded padlock probes, locally amplified and sequenced by repeated fluorescent staining and imaging cycles.
An \textit{in situ} sequencing dataset consists of five fluorescent channels for each sequencing cycle: one nuclei channel and four colour channels where the fluorescent signals belonging to the four bases of the genetic code (T, G, C, A) are imaged.
An additional general stain is used in the first cycle to detect all four bases in a single image used as reference.
Fluorescent signals appear as bright spots of 3-7px size in a noisy background caused by scattering light and autofluorescence. Moreover, fluorescent spots have a blurry appearance without any clear border due to the low signal-to-noise ratio caused by the diffraction limit of the microscope. Therefore, it is often difficult to distinguish real fluorescent signals from noise and background structures focusing the analysis on criteria based only on intensity values.
We previously published an image analysis pipeline for signal decoding of the barcoded padlock probes mapping targeted mRNAs with morphological and spatial information in cells and tissue \cite{pacureanu2014image}.
This previous approach detected fluorescent signals by applying a global threshold on intensities of the general stain image, followed by a watershed segmentation to resolve clusters. The resulting segmentation mask was then used to extract fluorescence intensities in the four colour channels for each sequencing cycle. Thereafter, barcodes were decoded selecting the base from the channel with the highest intensity for each sequencing cycle and finally filtered using a quality threshold based on intensity values.
As compared to visual assessment, many signals were missed using this approach.
In order to improve recall of the decoded barcode sequences we here present a pipeline that aims to be as inclusive as possible in the first processing steps. Thus, we delay the decision of which fluorescent signal candidates contribute to an expected sequence.
We use a Convolutional Neural Network (CNN) to extract self-learned features from signal candidates and use them as a probability prediction to describe how similar a signal candidate is compared to a true signal judged by visual examination. We thereafter feed signal candidates to a graphical model that resolves the sequences and provides the final barcodes. Finally, a quality measurement of the decoded sequences is assessed, making it possible for the user to set a threshold to achieve high recall or high accuracy.

\section{Method}
\subsection{Image Registration}
Before extracting the signals we align the images to compensate image misalignment among successive sequencing cycles.
We firstly apply a maximum intensity projection (MIP) of the general stain and the nuclei channel of the first sequencing cycle and we use the resulting image as a fixed reference. Second, for each sequencing cycle we combine the five channels (nuclei, T, G, C, A) by MIP and use them as moving images to perform a rigid registration to the reference using Elastix \cite{klein2010elastix}. Finally the same transformation matrices are applied to the four colour channels (T, G, C, A) of the respective sequencing cycle achieving their final registration.

\subsection{Signal Candidate Detection}
After image registration we extract signal candidates from each channel and each sequencing cycle performing a coarse detection including noise, since the graphical model resolves the sequences picking up only later high quality candidates to build up a barcode sequence.
This allow us to retrieve also the weakest signals that might be confused with noise and discarded using high threshold parameters.

\subsubsection{Normalization}
In order to be able to compare intensity values among different channels, we normalize the images scaling the intensities between the estimated background value (\textit{i.e.} mode of the image) and the brightest fluorescent signal values (\textit{i.e.} \(99^{th}\) percentile of intensity values):
\[{I\_norm_{ij}=\frac{I_{ij}-mode(I_{ij})}{P99(I_{ij})-mode(I_{ij})}}\]
where \(I_{ij}\) is the registered image of channel \(j\) at sequencing cycle \(i\) and \(P99(I_{ij})\) is its \(99^{th}\) percentile.

\subsubsection{Signal Candidate Extraction}
Before extracting the signal candidates, we first enhance bright spots and attenuate background contribution using Top-Hat filtering with a disk of radius 5 as structuring element.
Afterwards, we perform an h-maxima transformation \cite{soille2013morphological} in which all local maxima are found, but only those whose dynamic is higher than a given value \(h\) %, fixed to 0.1,
are extracted as signal candidates.
Furthermore, since some very bright signals present saturated intensity values, the h-maxima transform detects regions of local maxima consisting of several connected pixels. Thus, we consider the centroid coordinates  of these regions as signal candidate positions.
The result of this step is a mask, having the same size as the original image, with white pixels corresponding to the detected candidate positions.
This step is applied to the general stain channel image of the first sequencing cycle and to each colour channel of every sequencing cycle.

\subsubsection{Signals Merging}
Due to photo-bleaching and broad emission spectra, fluorescent signals can bleed-through to other channels. Thus, to filter multiple detections associated to the same fluorescent signal, the extracted signal candidates belonging to different masks are associated as follows:
\begin{enumerate}
    \item signal candidate coordinates of the general stain channel are extracted from the relative mask and taken as reference;
    \item for each couple of coordinates extracted we check, in each colour channel mask for each sequencing cycle, if there is a corresponding candidate having coordinates located in a 3x3 px window centred in the reference coordinates;
    \item if any corresponding candidate is found we retrieve the intensity value of the closest coordinate from the respective normalized image (otherwise intensities are extracted from each channel normalized image using the reference coordinates).
\end{enumerate}
After this step, for each associated candidate that belongs to the same fluorescent signal, we select the signal having the maximum fluorescence intensity.

\subsection{Signal Candidate Predictions}
For each signal candidate extracted in the previous step, we adopt a CNN based self-learning approach to determine the probability of being signal and noise. Using training data, the CNN learns the underlying discriminative features to predict the similarity between a signal candidate and a true signal.

The CNN architecture is composed of a convolutional layer with 200 filters of 5x5 convolutions, a fully connected layer of 128 neurons and a softmax layer with 2 output classes. Two dropout layers are used, as regularizers, before and after the fully connected layer to avoid overfitting.
The convolutional layer takes as input an array of 5x5 px windows. Each window is centred on a signal candidate location and contains the intensity values extracted from the relative normalized image.
The network was first trained on an annotated dataset from a different \textit{in situ} sequencing experiment. Using a new set of 200 annotated signals,
selected randomly across all color channels and sequencing cycles
we fine-tuned the pretrained network to compensate it for different experimental parameters and biological sources of the samples. The output is the probability predictions of a signal being true or noise.

\subsection{Graphical Model}
We finally resolve the sequences combining all signal candidates and probability predictions. The graphical model helps us to model the uncertainty of the data (probability that a detection is a true signal or not) in combination with a representation of the logical structure of the data itself (topological constraints applied to the graph that promote the formation of paths representing the decoded sequences).
Signal candidates are encoded in the graph as \textit{detection variables} represented as \(D\) nodes (Fig.~\ref{gm_fig}). Each detection variable is a boolean random variable that represents the nature of the detection: \textit{False} for false positive detection (noise), \textit{True} for true positive detection (signal). Relationships among signals candidates belonging to different sequencing cycles are encoded as \textit{transition variables}, represented as \(T\) nodes. Each transition variable connects a pair of detection variables representing signals candidates from any pair of different cycles whose distance is less then a given threshold \(d_{th}\). Therefore, transition variables are boolean random variables that assume \textit{True} values when a pair of true signal candidates belong to the same barcode sequence (\textit{i.e.} mRNA molecule).
We can formally describe our graphical model as a graph \(G=\{D,T,E,f,g\}\) where, \(D=\{D_1,D_2,\dots,D_n\}\) is the set of detection variables, \(T=\{T_1, T_2,\dots,T_m\}\) is the set of transition variables, \(E =\{e_1,e_2,\dots,e_z\}\) is the set of edges, \(f=\{f_1,f_2,\dots,f_n\}\) and \(g=\{g_1,g_2,\dots,g_m\}\) are cost functions respectively for candidate selection and aggregation among sequencing cycles defined as:
\begin{equation}
f_i(D_i)=\left\{%
    \begin{array}{lc}
         -log(p_0) & D_i=0 \\
         -log(p_1) & D_i=1
    \end{array}\right., \forall i \in [1,n]
\end{equation}
where $p_0$ and $p_1$ are probability predictions from the CNN of being respectively noise and signal; and
\begin{equation}
g_j(T_j)=\left\{%
    \begin{array}{lc}
         -log(\mu_t) & T_j=0 \\
         -log(1-\mu_t) & T_j=1
    \end{array}\right., \forall j \in [1,m]
\end{equation}
where $\mu_t$ is an affinity function that describes how close two signal candidates belonging to different cycles are related to each other in terms of distance and intensity value. Specifically:
\begin{equation}
    \mu_t = \frac{1}{(1+k_1 \cdot \Delta I)(1+k_2 \cdot d)}
\end{equation}
The affinity function $\mu_t$ is inversely proportional to the difference of intensity values between the pair of relative signal candidates $\Delta I$ and the euclidean distance $d$ between them. Two weighting parameters, $k_1$ and $k_2$, are used to modulate the contribution of $\Delta I$ and $d$.

The graph is solved minimizing the cost function $C$ defined as:
\begin{equation}
    C(D,T)=\sum_{i=1}^n f_i(D_i) + \sum_{j=1}^m g_j(T_j)
\end{equation}

% talk about CONSTRAINTS
However, because of the nature of the problem only certain configurations of the random variable states can encode a valid representation of a barcode sequence. Thus, in order to allow only a subset of the solution space, the following constraints are added to the graph:
\begin{enumerate}
    \item resolved sequences must have a length equal to the number of sequencing cycles,
    \item resolved sequences must be encoded by $D$ variables belonging to different sequencing cycles,
    \item each $T$ and $D$ can only encode a single barcode,
    \item if a $D$ variable is set to $False$, then all $T$ variables connected to it are set to $False$.
\end{enumerate}

\subsection{Quality of Decoded Barcodes}
Finally, we define a quality metric $Q_b$ for each decoded sequence $b$ encoded by the set $\{D_{b1},\dots,D_{bh}\} \subset D$, where $h$ is the number of sequencing cycles:
\begin{equation}
    Q_b = \frac{1}{\sum_{i=1}^h f_{bi}(D_{bi})}
\end{equation}

\section{Experiments and Results}
\begin{figure}[t]
\centering
\includegraphics[width=\columnwidth]{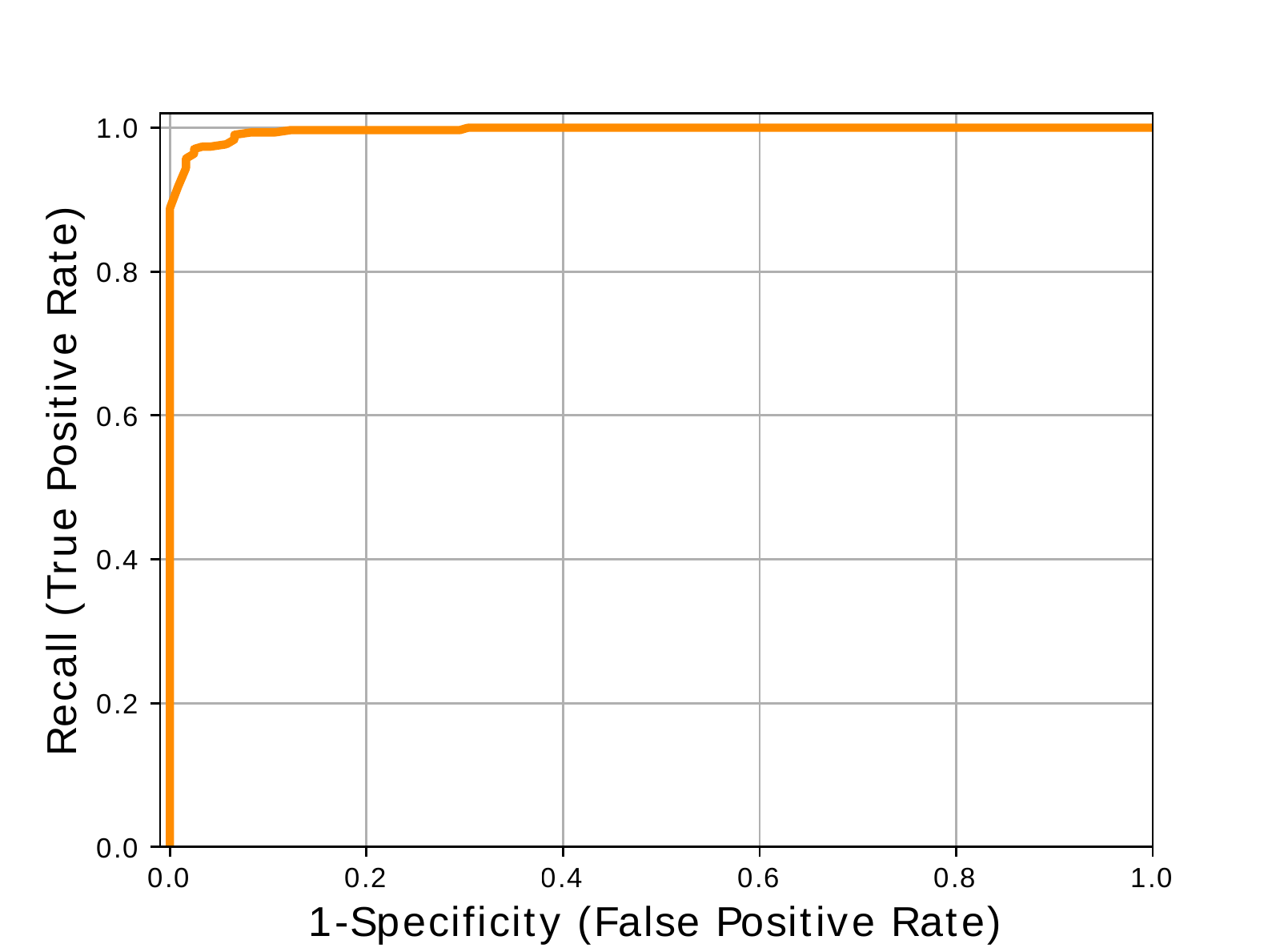}
\caption{ROC curve. The recall of decoded barcodes corresponding to the targeted sequences (\textit{true positive rate}) against $1-$specificity of decoded barcodes corresponding to unexpected sequences (\textit{false positive rate}), at decreasing quality thresholds.}
\label{TPFP}
\end{figure}

\begin{figure*}[t]
\centering
\includegraphics[width=0.935\textwidth]{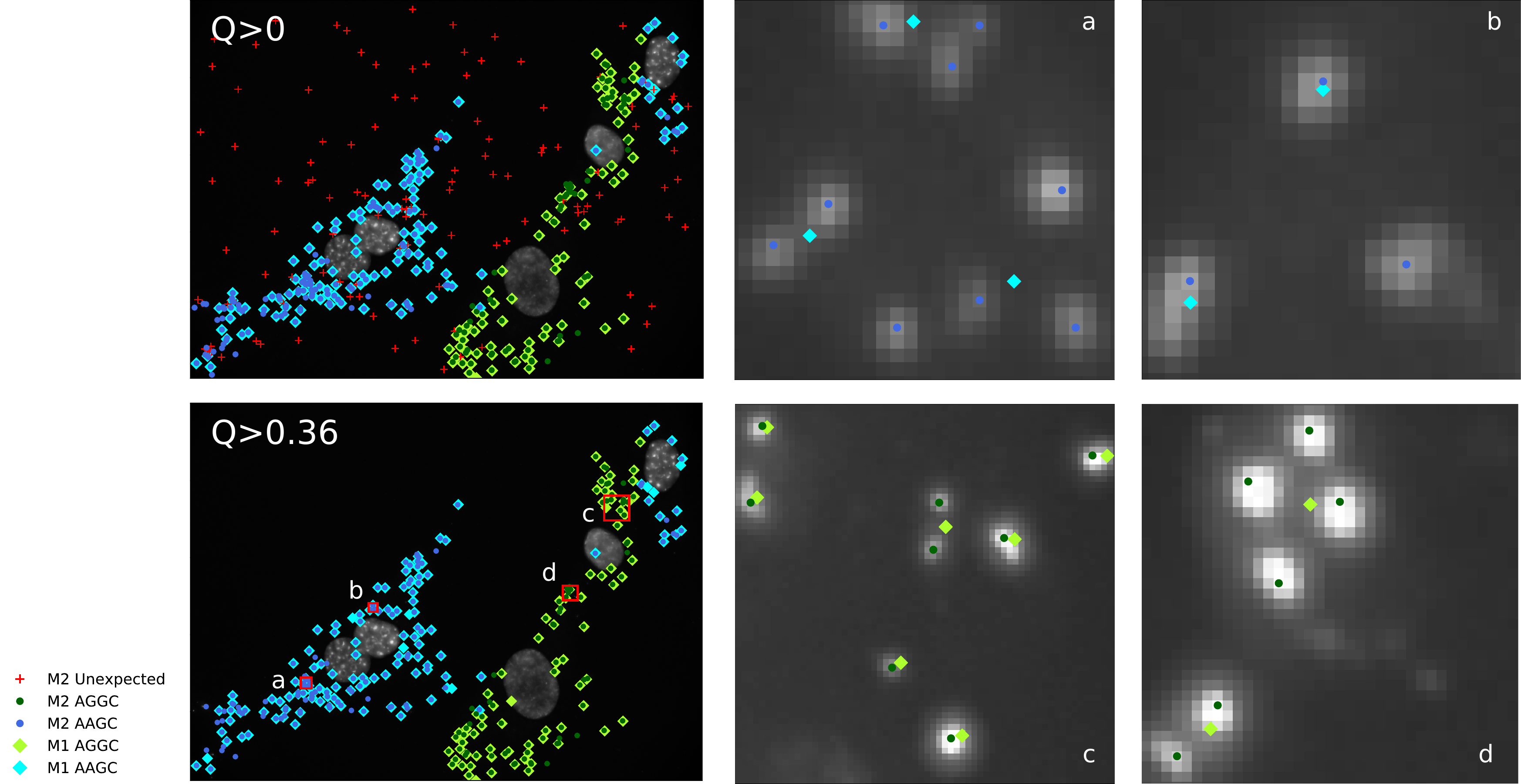}
\caption{Markers overlap on the general stain image of the two targeted barcodes (\textit{AGGC}, \textit{AAGC}) and unexpected barcodes, decoded by the state-of-the-art pipeline \textit{M1} \cite{pacureanu2014image} and the proposed method \textit{M2}.
The top-left image shows sequences decoded by \textit{M2} without any quality filtering.
The bottom-left image shows barcodes with a quality metric higher than a given threshold whose value has been set to filter out unexpected sequences. The right side of the figure shows the zooms of four different regions where overlapping markers of decoded barcode sequences with high quality are shown in detail. M2 often resolve signals that were merged by M1.}
\label{fig1}
\end{figure*}

\begin{figure*}[t]
\centering
\includegraphics[width=0.935\textwidth]{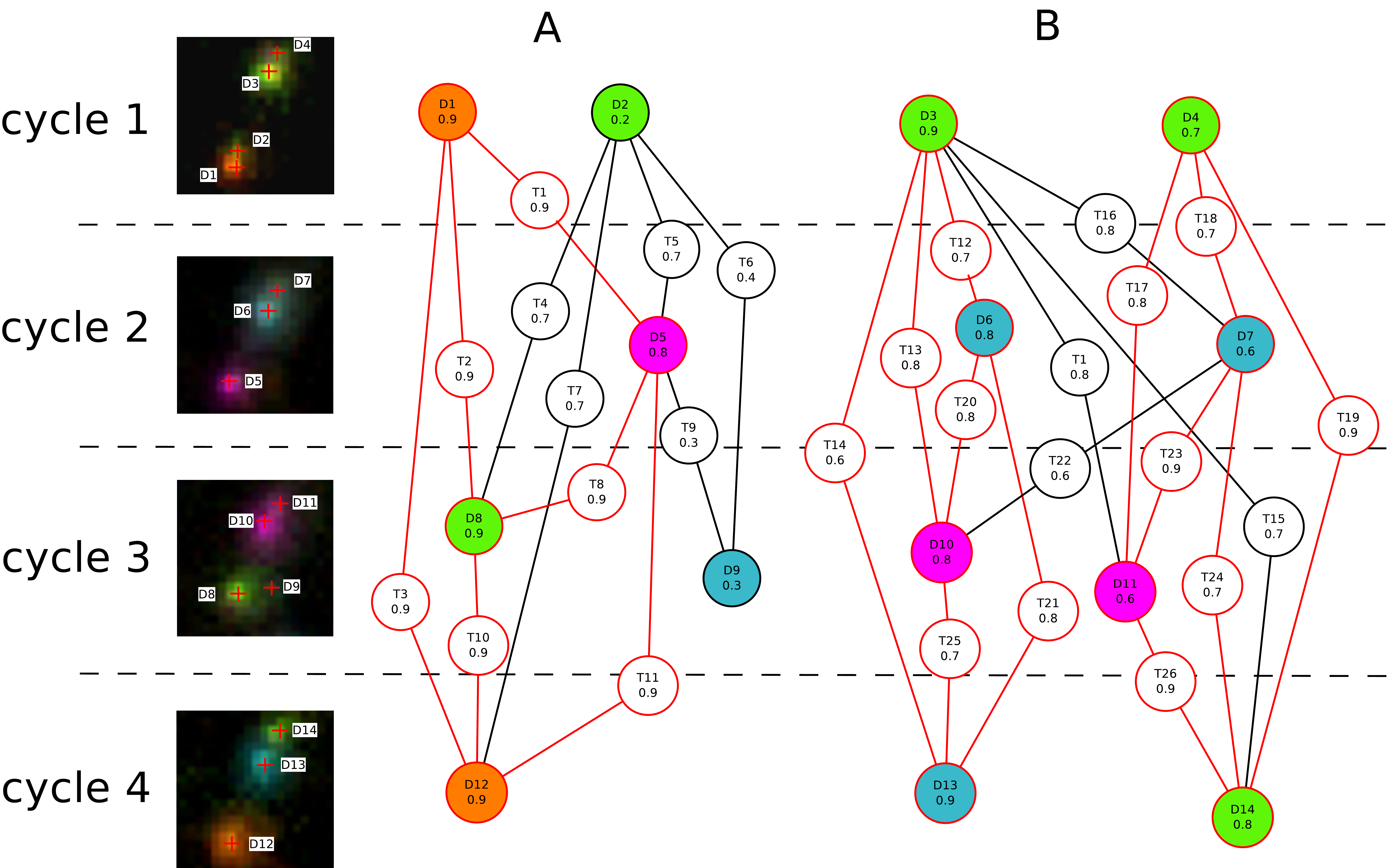}
\caption{Graphic representation of barcode sequences decoding example through the graphical model of three different mRNA molecules. On the left, labeled with numbers from 1 to 4, cut-out composite images of a small region from the four sequencing cycles. Fluorescent signals from different channels are shown with different colours (A in orange, C in green, G in magenta, and T in cyan). Each $D$ variable represents a signal detection marked with red crosses in each cut-out. For each pair of signal detections of two different cycles with distance smaller than \(d_{th}\), a $T$ variable is inserted and represented with white nodes in the graph. Each $D$ and $T$ nodes shows respectively the probability to be signal $p_1$ and the affinity value $\mu_t$. The graphical representation of the signal candidates present in the image cut-outs result in two independent connected components represented by the two graphs with labels A and B. Solving the graphs by energy minimization decodes \textit{AGCA}, \textit{CTGT}, \textit{CTGC} barcode sequences highlighted by the red paths in figure.
%Nodes that contribute to a barcode sequence are set to \textit{True}, while the others are set to \textit{False}.
}
\label{gm_fig}
\end{figure*}

To validate our method we performed sequence decoding on a previously published \textit{in situ} sequencing dataset \cite{ke2013situ}. The dataset consists of a co-culture of human and mouse cells where a four-base-long sequence of \textit{ACTB} mRNA has been sequenced. The sequence is identical in the two type of cells except for a variation in the second base that allow to differentiate between human and mouse mRNAs. The dataset has been acquired in 3D at different focal depths and then combined into a single image with MIP by the proprietary microscope software. We refer to \cite{ke2013situ} for further image acquisition details.
We evaluate the results in terms of recall in respect to the state-of-the-art image analysis pipeline \cite{pacureanu2014image} and in terms of precision evaluating the spatial correlation of the detected sequences in relation to the biological information. The analysis is performed running the pipeline with the parameters setting: $h=0.15$, $d_{th}=2$, $k_1=0.23$, and $k_2=0.1$. We consider resolved barcodes of the targeted gene (with sequence \textit{AGGC} and \textit{AAGC}) as \textit{true positive} (TP) decoded sequence. And sequences that differ from the expected barcodes as \textit{false positive} (FP) results.
We exclude sequences belonging to \textit{homopolymers} (e.g. barcodes consisting of a single letter, such as \textit{AAAA}, \textit{CCCC}, \textit{GGGG} and \textit{TTTT}) from the FP count. This kind of false detections arise from autofluorescent of biological structures that often appear very similar to real fluorescent signals, and are never used when designing barcodes for an \textit{in situ} sequencing experiment.
Comparison of results between the state-of-the-art and the proposed pipeline are shown in Fig.~\ref{fig1}. The proposed pipeline decodes $304$ TP and $122$ FP sequences prior to quality thresholding. %%%%%124
After evaluating the receiver operating characteristics (ROC) (Fig.~\ref{TPFP}), we selected a quality threshold to exclude unexpected sequences that generally have a low quality metric since they often come from noise and false detections. Setting the quality threshold high enough to exclude all false positive results lead to $270$ TP signals compared to $213$ barcodes correctly identified by the state-of-the-art pipeline.

\section{Conclusion}
To conclude, the proposed signal detection and decoding approach increases signal recall by $27\%$ while maintaining detection sensitivity. Based on a visual comparison of the proposed approach and the state-of-the-art, it is clear that clustered signals are resolved with higher accuracy. Increased recall will have impact on all subsequent spatial statistics of tissue heterogeneity approached by \textit{is situ} sequencing.

\section*{Acknowledgment}
This research is part of the TissueMaps project, funded by the European Research council via ERC Consolidator grant 682810 to C. W{\"a}hlby. The data was kindly provided by M. Nilsson and his research group at SciLifeLab Stockholm.

%\vspace{2mm} % this command adds a little vertical space to fine-tune
             % the column balancing. LaTeX has no mechanism to balance
             % columns.

% trigger a \newpage just before the given reference
% number - used to balance the columns on the last page
% adjust value as needed - may need to be readjusted if
% the document is modified later
%\IEEEtriggeratref{8}
% The "triggered" command can be changed if desired:
%\IEEEtriggercmd{\enlargethispage{-5in}}

% references section

\bibliographystyle{IEEEtran}
\bibliography{sample}

% that's all folks
\end{document}